\documentclass[11pt,english]{article}
\usepackage{amssymb, amsmath}
\usepackage{multicol, url, color}
\usepackage{amsmath, amssymb}
\usepackage[mathscr]{eucal}
\usepackage{graphicx}
\usepackage{amsfonts}
\usepackage{babel}
\usepackage{framed}
\newcommand{\bfi}{\bfseries\itshape}
\newcommand{\rem}[1]{}
\newcommand{\remfigure}[1]{}

\makeatletter
\@addtoreset{equation}{section}

\makeatother

\def\contract{\makebox[1.2em][c]{\mbox{\rule{.6em}
{.01truein}\rule{.01truein}{.6em}}}}

\newtheorem{theorem}{Theorem}

\newtheorem{assumption}[theorem]{Assumption}

\newtheorem{remark}[theorem]{Remark}

\numberwithin{theorem}{section}

\def\0{{\bf 0}}

\begin{document}
\pagestyle{myheadings}
\markright{Holm, Putkaradze, Tronci \hfil Double bracket dissipation in kinetic theory \hfil  \today}
%\Large
\title{
%
%Singular solutions for geodesic flows of Vlasov moments
Double bracket dissipation in kinetic theory\\ for particles with anisotropic interactions
%with applications to fluids and accelerator beams
}
\author{\vspace{2mm}
D. D. Holm$^{1,\,2}$, V. Putkaradze$^{3,\,4}$ and C. Tronci$^{5}\!$\\
%\footnote{
%$\,$also at {\it TERA Foundation for Oncological Hadrontherapy, 11 Via Puccini, %Novara 28100, Italy}
%}
\\
{\small \!\!\!$^1$ \it Department of Mathematics, Imperial College London, London SW7 2AZ, UK}\\
{\small \!\!\!$^2$ \it Institute of of Mathematical Sciences, Imperial College London, London SW7 2AZ, UK}\\
{\small \!\!\!$^3$ \it Department of Mathematics, Colorado State University,
Fort Collins, CO 80523 USA}\\
{\small \!\!\!$^4$ \it Department of Mechanical Engineering, University of New Mexico,
Albuquerque NM 87131 USA}\\
{\small \!\!\!$^5$\,\it Section de Math\'ematiques, \'Ecole Polytechnique
F\'ed\'erale de Lausanne. Switzerland}
\\ \\
}
\date{\today}

\maketitle

\begin{abstract}\noindent
We derive equations of motion for the dynamics of anisotropic particles directly from the dissipative Vlasov kinetic equations, with the dissipation given by the double bracket approach
(Double Bracket Vlasov, or DBV). The moments of the DBV equation lead to a nonlocal form of Darcy's law for the mass density. Next, kinetic equations for particles with anisotropic interaction are considered and also cast into the DBV form. The moment dynamics for these double bracket kinetic equations is expressed as Lie-Darcy continuum equations for densities of mass and orientation. We also show how to obtain a Smoluchowski model from a cold plasma-like moment closure of  DBV.
Thus, the double bracket kinetic framework  serves as a unifying method for deriving different types of dynamics, from density--orientation to Smoluchowski equations. Extensions for more general physical systems are also discussed.
\end{abstract}

%\newpage

%\tableofcontents

%\newpage

%%%%%%%%%%%%%%%%
\section{Introduction}
\subsection{Geometric models of dissipation in physical systems}

This paper explains
how the geometry of double-bracket dissipation makes its way from the microscopic (kinetic theory) level to the macroscopic (continuum) level in the process of taking moments of the Vlasov probability distribution function, when the particles in the microscopic description carry an internal variable that is orientation dependent.  Without orientation dependence, the moment equations derived here yield a nonlocal variant of the famous Darcy law \cite{Darcy1856}. When orientation is included, the resulting {\bfi Lie-Darcy} moment equations identify the macroscopic parameters of the continuum description and govern their evolution.

In previous work, Gibbons, Holm and Kupershmidt \cite{GiHoKu1982,GiHoKu1983} (abbreviated GHK) showed that the process of taking moments of Vlasov's equation for such particles is a Poisson map.  GHK used this property to derive the equations of {\bfi chromohydrodynamics}. These are the equations of a fluid plasma consisting of particles carrying Yang-Mills charges and interacting self-consistently via a Yang-Mills field. The GHK Poisson map for chromohydrodynamics was an extension of the Kupershmidt-Manin {\bfi (KM)} bracket \cite{KuMa1978} for the moments in the integrable hierarchy of  long-wave equations due to Benney \cite{Be1973}. Gibbons \cite{Gi1981} had noticed that the KM Poisson map for the Benney equations would also apply to the moments of a Vlasov (collisionless) plasma  \cite{Vlasov1961} for particles carrying an ordinary electric charge. GHK extended the KM bracket to allow the particles to carry orientation (or Yang-Mills) properties. GHK considered only Hamiltonian motion. They did not consider the corresponding double-bracket Poisson structure of dissipation. That is the subject of the present work.

\subsubsection{History of double-bracket dissipation}
Bloch, Krishnaprasad, Marsden and Ratiu (\cite{BlKrMaRa1996} abbreviated BKMR) observed that linear dissipative terms of the standard Rayleigh dissipation type are inappropriate for dynamical systems undergoing coadjoint motion. Such systems are expressed on the duals of Lie algebras and they commonly arise from variational principles defined on tangent spaces of Lie groups. A well known example of coadjoint motion is provided by Euler's equations for an ideal incompressible fluid \cite{Ar1966}. Not unexpectedly, adding linear viscous dissipation to create the Navier-Stokes equations breaks the coadjoint nature of the ideal flow.  Of course, ordinary viscosity does not suffice to describe dissipation in the presence of orientation-dependent particle interactions.

Restriction to coadjoint orbits requires nonlinear dissipation, whose gradient structure differs from the Rayleigh dissipation approach leading to Navier-Stokes viscosity.  As a familiar example on which to build their paradigm, BKMR emphasized a form of energy dissipation (Gilbert dissipation \cite{Gilbert1955}) arising in models of ferromagnetic spin systems that preserves the magnitude of angular momentum. In the context of Euler-Poincar\'e or Lie-Poisson systems, this means that coadjoint orbits remain invariant, but the energy decreases along the orbits. BKMR discovered that their geometric construction of the nonlinear dissipative terms summoned  the double bracket equation of Brockett \cite{Br1988, Br1993}. In fact, the double bracket form is well adapted to the study of dissipative motion on Lie groups since it was originally constructed as a gradient system \cite{Bloch1990,BlBrRa1992,Br1994,BlFlRa1996}.

While a single Poisson bracket operation is bilinear and antisymmetric, a double bracket operation is a symmetric operation. Symmetric brackets for dissipative systems, particularly for fluids and plasmas, were considered previously by Kaufman \cite{Ka1984, Ka1985}, Grmela \cite{Gr1984, Gr1993a, Gr1993b}, Morrison \cite{Mo1984, Mo1986}, and Turski and Kaufman \cite{TuKa1987}. The dissipative brackets introduced in BKMR were particularly motivated by the double bracket operations introduced in Vallis, Carnevale, and Young \cite{VaCaYo1989} for incompressible fluid flows.

\subsubsection{Selective decay hypothesis}
One of the motivations for Vallis et al.  \cite{VaCaYo1989} was the {\bfi selective decay hypothesis}, which arose in turbulence research \cite{MaMo1980} and is consistent with the preservation of coadjoint orbits.
According to the selective decay hypothesis, energy in strongly nonequilibrium statistical systems tends to decay much faster than certain other ideally conserved properties. In particular, energy decays much faster in such systems than those ``kinematic'' or ``geometric'' properties that would have been preserved in the ideal nondissipative limit {\it independently of the choice of the Hamiltonian}. Examples are the Casimir functions for the Lie-Poisson formulations of various ideal fluid models \cite{HoMaRaWe1985}.

The selective decay hypothesis was inspired by a famous example; namely, that enstrophy decays much more slowly than kinetic energy in 2D incompressible fluid turbulence. Kraichnan \cite{Kr1967} showed that the decay of kinetic energy under the preservation of enstrophy causes dramatic effects in 2D turbulence. Namely, it causes the well known ``inverse cascade'' of kinetic  energy to {\it larger} scales, rather than the usual ``forward cascade'' of energy to smaller scales that is observed in 3D turbulence.  In 2D ideal incompressible fluid flow the enstrophy (the $L^2$ norm of the vorticity) is preserved on coadjoint orbits. That is, enstrophy is a Casimir of the Lie-Poisson bracket in the Hamiltonian formulation of the 2D Euler fluid equations. Vallis et al. \cite{VaCaYo1989} chose a form of dissipation that was expressible as a double Lie-Poisson bracket. This choice of dissipation preserved the enstrophy and thereby enforced the selective decay hypothesis for all 2D incompressible fluid solutions, laminar as well as turbulent.

Once its dramatic effects were  recognized in 2D turbulence,  selective decay was posited as a governing mechanism in other systems, particularly in statistical behavior of fluid systems with high variability. For example, the slow decay of magnetic helicity was popularly invoked as a possible means of obtaining magnetically confined plasmas \cite{Ta1986}.  Likewise, in geophysical fluid flows, the slow decay of potential vorticity (PV) relative to kinetic energy strongly influences the dynamics of weather and climate patterns much as in the inverse cascade tendency in 2D turbulence. The use of selective decay ideas for PV thinking in meteorology and atmospheric science has become standard practice since the fundamental work in \cite{HoMcRo1985, Yo1987}.

\subsubsection{Kandrup and the double bracket for astrophysical instabilities}
A form of selective decay based on double-bracket dissipation was proposed in astrophysics by Kandrup \cite{Ka1991} for the purpose of modeling gravitational radiation of energy in stars. In this case, the double-bracket dissipation produced rapidly growing instabilities that again had dramatic effects on the solution. The form of double-bracket dissipation proposed in Kandrup \cite{Ka1991} is a strong motivation for the present work and it also played a central role in the study of instabilities in BKMR.

The double bracket idea mentioned earlier in the context of magnetization dynamics \cite{Gilbert1955} implements dissipation by the sequential application of  two Poisson bracket operations. The dissipative Vlasov equation in \cite{Ka1991} is written as:
\begin{equation}
\frac{\partial f}{\partial t}+ \left\{f,\,\frac{\delta H}{\delta f}\right\}
=
\alpha \left\{f,\left\{f,\,\frac{\delta H}{\delta f}\right\} \right\}
\,,
\label{Kandrup-dbrkt}
\end{equation}
where $f(q,p,t)$ is the probability distribution function for a
single particle in phase-space, $\alpha>0$ is a positive constant,
$H$ is the Vlasov Hamiltonian (depending on the system under
consideration) and $\{ \cdot \, , \, \cdot \}$ is the canonical
Poisson bracket in $(q,p)$. Here we use standard notation in kinetic
theory: in the general case, $q$ is a point on a Riemannian manifold
$\mathcal{Q}$, while $(q,p)\in T^*\mathcal{Q}$ is a point in the
cotangent bundle. In common situations one considers
$\mathcal{Q}=\Bbb{R}^k$, so that $T^*\mathcal{Q}=\Bbb{R}^{2k}$. In
this paper we shall consider a one dimensional physical space
($\mathcal{Q}=\Bbb{R}$) to simplify the treatment.

When $\alpha\to0$ in \eqref{Kandrup-dbrkt}, this equation reduces the Vlasov equation for collisionless plasmas. For $\alpha>0$, this is the double bracket dissipation approach for the Vlasov-Poisson equation.
This nonlinear double bracket approach for introducing dissipation into the Vlasov equation differs from the standard Fokker-Planck linear diffusive approach \cite{Fokker-Plank1931}, which adds dissipation on the right hand side of equation (\ref{Kandrup-dbrkt}) as a linear term given by the Laplace operator in the momentum coordinate $\Delta_p f$.

\subsubsection{The double bracket and Riemannian geometry}
 An interesting feature of the double bracket formulation is that it  leads via a variational
approach to a symmetric Leibnitz bracket that in turn yields a
metric tensor and an associated Riemannian (rather than symplectic)
geometry for the solutions.  The variational approach thus preserves
the nature of the evolution of Vlasov phase space density, by
coadjoint motion under the action of the canonical transformations
on phase space densities.

 As Otto \cite{Ot2001} explained, the Riemannian geometry
of dissipation may be revealed by understanding how it emerges from a variational principle. Here, we follow the variational approach and consider a generalization of the double bracket structure in equation (\ref{Kandrup-dbrkt}) that recovers previous cases for particular
choices of modeling quantities introduced in \cite{HoPuTr2007-CR},
\begin{equation}
\frac{\partial f}{\partial t}+ \left\{f,\,\frac{\delta H}{\delta f}\right\}
=
 \left\{f,\,\left\{\mu[f]\,,\,\frac{\delta E}{\delta f}\right\} \right\}
\,.
\label{Vlasov-diss}
\end{equation}

The important feature of this double-bracket approach to dissipation  (which is necessarily nonlinear) for the Vlasov equation is that this form of the dissipation preserves the most basic property of Vlasov dynamics. Namely, {\bfi the dissipation is introduced as a canonical transformation}.
This means that the solution is still an {\bfi invariant probability distribution and satisfies Liouville's theorem}, even though its dynamics do not conserve either the Hamiltonian $H$ or the energy $E$.
Eq. (\ref{Vlasov-diss})  generalizes the double bracket operation in Eq. (\ref{Kandrup-dbrkt}) and reduces to it when the Hamiltonian $H$ is identical to the dissipated energy $E$ and the mobility $\mu[f]=\alpha f$ is proportional to the Vlasov distribution function $f$ for a positive constant $\alpha>0$. As shown in \cite{HoPuTr2007-CR}  the  generalization (\ref{Vlasov-diss}) has important effects on the types of solutions that are available to this equation.
Indeed, the form (\ref{Vlasov-diss}) of the Vlasov equation with dissipation allows for more general mobilities than those considered in \cite{BlKrMaRa1996,Ka1991,Ka1984,Mo1984}.
For example, one may choose mobility in the form $\mu[f]=K*f$, where  the $(\,*\,)$ operation is a convolution in phase space with an appropriate kernel $K$. In \cite{HoPuTr2007} the smoothing operation in the definition of $\mu[f]$ introduces a fundamental action scale (the area in phase space associated with the kernel $K$) into the dissipation mechanism. Accordingly, the dissipation depends on phase-space averaged quantities, rather than local pointwise values.

This smoothing also has the fundamental advantage of endowing (\ref{Vlasov-diss}) with the {\bfi one-particle solution as its singular solution}.
The generalization Eq. (\ref{Vlasov-diss}) may also be justified by using  thermodynamic and geometric arguments \cite{HoPuTr2007}.
Section \ref{sec:Dissvlasov} shows that this generalization leads from the microscopic kinetic level to the classic Darcy law (velocity being proportional to force) in the continuum description.

\subsection{Our goal and approach}
The goal of the present work is to determine the macroscopic
implications of introducing nonlinear double-bracket dissipation at
the microscopic level, so as to respect the coadjoint orbits of
canonical transformations for dynamics that depends upon particle
orientation. Our approach introduces this orientation dependence
into the microscopic description by augmenting the canonical Poisson
bracket in position $q$ and momentum $p$ so as to include the
Lie-Poisson part for orientation $g$ taking values in the dual
$\mathfrak{g}^*$ of the Lie algebra $\mathfrak{g}$, with eventually
$\mathfrak{g}=\mathfrak{so}(3)$ for physical orientation. Here we
denote by $\mathcal{G}$ the Lie group underlying the particle
symmetry properties, while the corresponding Lie algebra
$\mathfrak{g}=T_e\mathcal{G}$ and its dual
$\mathfrak{g}^*=T_e^*\mathcal{G}$ are denoted by gothic fonts.
Although it is unusual to denote a point in $\mathfrak{g}^*$ by $g$,
we shall keep this notation to avoid confusion with the Greek
indexes used below.

The identification of the particle orientation $g$ with an element
in $\mathfrak{so}^*(3)$ is consistent with ordinary theories in
condensed matter physics, such as the theory of sin glasses. In this
context, the spin angular momentum variable $s\in\mathfrak{so}^*(3)$
possesses the typical Poisson bracket
$\{s_\mu,s_\nu\}=\varepsilon_{\mu\nu\lambda\,} s_\lambda$
\cite{Dzyaloshinskii-Volovik1980}, where the Levi-Civita symbol
$\varepsilon$ denotes the total antisymmetric tensor. In more
general situations when the rotational symmetry is completely
broken, one also needs to consider the particle rotational state
$\Theta\in SO(3)$ as an extra Lie group coordinate in the augmented
phase-space. This is relevant, for example, when one wants to
consider frustration effects. However, here we shall not deal with
symmetry breaking and we shall simply assume that the orientational
degree of freedom can be identified with the momentum variable
$g\in\mathfrak{so}^*(3)$, or in more generality
$g\in\mathfrak{g}^*$.

From the considerations above, one concludes that the introduction
of the orientational degree of freedom $g$ leads to a Vlasov
distribution $f(q,p,g,t)$ on the augmented phase space
$T^*\mathcal{Q}\times\mathfrak{g}^*$, whose {\bfi total Poisson
bracket} is written in the GHK form as

\begin{equation}
\Big\{f,h\Big\}_{\!1}:=\,
\Big\{f,h\Big\}+\left\langle g,\,\left[\frac{\partial f}{\partial g},\frac{\partial
h }{\partial g}\right]\right\rangle
\,,
\label{PB1}
\end{equation}
where $[\,\cdot\,,\,\cdot\, ]:\, \mathfrak{g}\times\mathfrak{g}\to
\mathfrak{g}$ is the Lie algebra bracket and
$\langle\,\cdot\,,\,\cdot\, \rangle:\,
\mathfrak{g}^*\times\mathfrak{g}\to \mathbb{R}$ is the pairing
between the Lie algebra $\mathfrak{g}$ and its dual
$\mathfrak{g}^*$. For rotations, $\mathfrak{g}=\mathfrak{so}(3)$ and
the bracket $[\,\cdot\,,\,\cdot\, ]$ becomes the cross product of
vectors in $\mathbb{R}^3$. Correspondingly, the pairing
$\langle\,\cdot\,,\,\cdot\, \rangle$ becomes the dot product of
vectors in $\mathbb{R}^3$. We consider the double-bracket dynamics
of $f(q,p,g,t)$ resulting from replacing the canonical Poisson
brackets in Eq. (\ref{Vlasov-diss}) by the direct sum of canonical
and Lie-Poisson brackets $\{\,\cdot\,,\,\cdot\,\}_1$ in Eq.
(\ref{PB1}). We then take moments of the resulting dynamics of
$f(q,p,g,t)$ with respect to momentum $p$ and orientation $g$, to
obtain the dynamics of the macroscopic description. The moments with
respect to momentum $p$ alone provide an intermediate set of
dynamical equations for the $p-$moments,
\begin{equation}\label{Smoluchowski-moments}
A_n(q,g,t):=\!\int p^n\,f(q,p,g,t)\,dp \,\,\,\quad\text{with}\quad
g\in\mathfrak{g}^*
 \,.
\end{equation}
(In higher dimensions one replaces the power $p^n$ with the tensor
power ${\bf p}^{\otimes\,n}$, also denoted by $\otimes^n{\bf p}$.)
These intermediate dynamics are reminiscent of the Smoluchowski
equation for the probability $A_0(q,g,t)$. We shall show how the
derivation of usual Smoluchowski-type equations from double-bracket
arguments may require special care depending on the chosen moment
closure. In particular, we formulate a new Smoluchowski equation,
which encodes the Landau-Lifshitz-Gilbert dissipation for
ferromagnetic particles.

We also find a closed set of continuum equations for the $(p,g)$ moments
of the type
\begin{equation}\label{Darcy-moments}
A_{n,\nu}(q,t)\,:=\int p^n\,g^\nu\,f(q,p,g,t)\,dp\,dg\,.
\end{equation}
where $g^\nu$ denotes the tensor power $g^\nu=g\otimes\dots\otimes g$, $\nu$ times.
The final closure provides the macroscopic continuum dynamics for
the set of moments of the double-bracket kinetic equations
(\ref{Vlasov-diss}) under the replacement
$\{\,\cdot\,,\,\cdot\,\}\to \{\,\cdot\,,\,\cdot\,\}_1$. This
macroscopic continuum closure inherits the geometric properties of
the double bracket, because the process of taking these moments is a
Poisson map, as observed in GHK
\cite{GiHoKu1982,GiHoKu1983,GiHoTr2008}.

\subsection{Mathematical framework for geometric dissipation}
As explained in \cite{Ot2001} dissipation of energy $E$ may naturally summon an appropriate metric tensor.
In previous work Holm and Putkaradze \cite{HoPu2007, HoPuTr2007} introduced a distance between functions on a vector space $V$ that is acted on by a Lie algebra $\mathfrak{X}$ of vector fields.
In particular they showed that that for any two functionals  $F[ \kappa],G[\kappa]$ of a geometric tensor quantity $\kappa\in V$ a distance between them may be defined via the  Riemannian metric,
\begin{equation}
g_\kappa \left(  F \, , \,  E \right) =
\left\langle
\Big(\mu[\kappa] \,\diamond\, \frac{\delta F}{\delta f}\Big)
,\,
\Big( \kappa\,\diamond\,\frac{\delta E}{\delta f}\,\Big)^\sharp
\right\rangle_{\mathfrak{X}^*\times\mathfrak{X}}
  \,.
\label{FGdist}
\end{equation}
Here $\langle\,\cdot\,,\,\cdot\, \rangle$ denotes the $L^2$ pairing
of vector fields in $\mathfrak{X}$ with their dual one-form
densities in $\mathfrak{X}^*$,  sharp $(\,\cdot\,)^\sharp$ raises
the vector index from covariant to contravariant and $\mu[\kappa]$
is the \emph{mobility functional}. The mobility $\mu[\kappa]$ is
assumed to satisfy the requirements for (\ref{FGdist}) to be
positive definite and symmetric, as discussed in \cite{HoPuTr2007}.
The diamond operation $(\diamond)$ in equation (\ref{FGdist}) is the
dual of the Lie algebra action, defined as follows. Let a vector
field $\xi\in\mathfrak{X}$ act on a vector space $V$ by Lie
derivation, so that the Lie algebra action of $\xi$ on any element
$\kappa\in V$ is given by the Lie derivative, $
\xi\!\cdot\kappa=\pounds_{\!\xi}\,\kappa $. The operation dual to
the Lie derivative is denoted by $\diamond$ and  defined in terms of
the $L^2$ pairings between $\mathfrak{X}$ and $\mathfrak{X}^*$ and
between $V$ and $V^*$ as
\begin{equation}\label{diamond-def}
\Big\langle \zeta\diamond\kappa,\xi
\Big\rangle_{\mathfrak{X}^*\times\mathfrak{X}}
:=
\Big\langle \zeta,-\pounds_{\!\xi}\,\kappa
\Big\rangle_{V^*\times V}
\,.
\end{equation}
Given the metric (\ref{FGdist}) and a dissipated energy functional $E[\kappa]$, the time evolution of \emph{arbitrary} functional $F[\kappa]$ is given by \cite{HoPu2007,HoPuTr2007} as
\begin{equation}
\dot{F}=\{\!\{\,F\,,\,E\,\}\!\} [\kappa] :=-\,g_\kappa \left( F, E
\right) \,, \label{bracket}
\end{equation}
which specifies the dynamics of any functional $F[\kappa]$, given
the  the energy dependence  $E[\kappa]$. The bracket
$\{\!\{\,F\,,\,E\,\}\!\}$ is shown to satisfy the Leibnitz
product-rule property for a suitable class of mobility functionals
$\mu[\kappa]$ in \cite{HoPu2007,HoPuTr2007}. Eq. (\ref{bracket}) and
positivity of $g_\kappa( E,E)$ imply that the energy $E$ decays in
time until it eventually reaches a critical point, $\delta E/\delta
\kappa=0$. For densities (dual to functions in the $L^2$ pairing),
the Lie derivative is the divergence:
$\pounds_\xi\,\kappa=\operatorname{div}\!\left(\kappa\,\xi\right)$
and its dual operation is (minus) the gradient: $\kappa\,
\diamond\,\zeta=-\kappa\,\nabla\zeta$. Thus, for densities the
symbol diamond $(\,\diamond\,)$ is replaced by gradient
$(\,\nabla\,)$ in the metric defined in (\ref{bracket}).

The definition of the dissipative bracket in Eq. (\ref{bracket}) for
arbitrary functionals $\{\!\{\,F\,,\,E\,\}\!\} [\kappa]$ is the
basis for our present considerations of dissipation in kinetic
equations. In this paper we will extend the geometric dissipation
(\ref{bracket}) to the symplectic case by defining the star
$(\,\star\,)$ operator, which is the analogue of diamond
$(\,\diamond\,)$ for symplectic spaces.

%\comment{CT: plan taken out}

\section{Dissipation in the Vlasov equation and the moment hierarchy} \label{sec:Dissvlasov}

\subsection{Dissipative bracket for the Vlasov equation}
The dissipative term in equation (\ref{Vlasov-diss}) is found by
considering the symplectic algebra action of a Hamiltonian vector
field $X_h$ associated with a Hamiltonian function $h$. This action
is given on a phase space density $f$ through the canonical Poisson
bracket $\{ \cdot \, , \, \cdot \}$ as follows \cite{MaWe}: $
\pounds_{\!X_h}\,f=\{f,h\}=:h\cdot f$.
One can check
that the dual diamond operator in \eqref{diamond-def} (denoted by $\star$ for the symplectic Lie algebra action above) is still a Poisson bracket \cite{HoPuTr2007}, so that $g\star f=\{g,f\}$.
Thus, we introduce the dissipative bracket (\ref{bracket})
\begin{equation}
\{\!\{\,E\,,\,F\,\}\!\}
=
-\,\Bigg\langle \left\{\,\mu[f]\,, \frac{\delta E}{\delta f}\right\}^\sharp,\, \left\{\,f\,,\frac{\delta F}{\delta f}\right\} \Bigg\rangle
=
\Bigg\langle \left\{f,\left\{\mu[f]\,, \frac{\delta E}{\delta f}\right\}^\sharp\right\},\,\frac{\delta F}{\delta f} \Bigg\rangle
\label{bracket-star}
\end{equation}
which gives the dissipative term in equation (\ref{Vlasov-diss}). Here, the sharp $\sharp$ operator is a suitably chosen mapping from the (dual) space of distributions to the Lie algebra of phase space functions, so the pairing can be taken.
Usually, we take the pairing to be a simple $L^2$ pairing, in which case the sharp notation is unnecessary, as in the previously postulated form of the equations (\ref{Vlasov-diss}).  The sharp notation in (\ref{bracket-star}) is used to indicate that the pairing may  involve a metric.

\begin{assumption}
\label{assumption-finitedim}
Everywhere in this paper, we assume that the Lie algebra $\mathfrak{g}$ of orientation motions is finite dimensional. Of particular interest to application is the motion of magnetized particles, described by $SO(3)$ group. For these motions, $\mathfrak{g} = \mathfrak{so}(3) \simeq \mathbb{R}^3$, so the integrals with respect to $g \in \mathfrak{g}$
are simply three-dimensional integrals.
\end{assumption}
Extensions of our theory could be computed for an arbitrary \emph{finite-dimensional} Lie algebras, describing, for example, particles with a finite number of joints. However, in this paper we will stay away from infinite-dimensional Lie algebras, that may arise, for example, for the case of flexible filaments. Even though the formal extension of our double bracket dissipation theory is possible even in that case, the mathematical difficulties of defining integrals and corresponding measures are beyond the scope of this paper. On the other hand, the canonical variables momenta $p$ and coordinates $q$ describe the motion of the center of mass of the particles and thus the integrals with respect to $p$ and $q$ are always finite-dimensional. In order to elaborate further, we introduce the following
\begin{assumption}
\label{assumption-existence}
Everywhere in the paper we assume that all the \emph{finite-dimensional} integrals defining the moments (\ref{Darcy-moments}) exist. Formally, this can be achieved, for example, by requiring that the distribution function $f$ is smooth and decays to zero sufficiently fast whenever $p$, $q$ or $g$ tend to $\infty$ in some properly defined norm. While for compact configuration manifolds requirements on $q$ integration convergence are automatically satisfied, the convergence requirements still need to be enforced for $p$ and $g$.
\end{assumption}

An interesting point about this form of double-bracket dissipation in kinetic equations is that it leads rather naturally to another widely used dissipative equation; namely, the venerable {\bfi Darcy law} \cite{Darcy1856}. In particular, we will show in this section how one recovers the nonlocal version of Darcy's law introduced in \cite{HoPu2005,HoPu2006}.
This  could be accomplished by integrating the purely dissipative
kinetic equation
\begin{equation}\label{diss_vlasov}
\frac{\partial f}{\partial t}
\,=\,
\left\{f\,,\,\left\{\mu[f]\,,\,\frac{\delta E}{\delta f}\,\right\}
\right\}
\,,
\end{equation}
with respect to the momentum coordinate $p$ (here we have dropped the sharp notation for simplicity). We shall call this equation \emph{Double Bracket Vlasov} equation (DBV). Note however, that this DBV equation is valid for isotropic particles only. More general version of DBV equation including the anisotropy will be derived below in (\ref{Vlasov-diss1}).
In order to derive Darcy's law for particle motion, we apply   the geometric dissipative bracket (\ref{bracket}) to the equations of the moment hierarchy.

\subsection{The Kupershmidt-Manin bracket for 1D moment dynamics}\label{Kup-Man}
As a general result \cite{Gi1981}, the equations for the moments of
the Vlasov dynamics (Eq. (\ref{Kandrup-dbrkt}) with $\alpha\to0$)
form a Lie-Poisson system under the Kupershmidt-Manin (KM) bracket
\cite{KuMa1978}. The $n$-th moment is defined as $A_n(q,t):=\int
p^n\, f(q,p,t)\, dp$. These quantities have a geometric
interpretation in terms of (covariant) tensor densities
\cite{GiHoTr2005,GiHoTr2007,GiHoTr2008} which can be seen by
re-writing the moments as
$A_n=$\makebox{$\int_p \otimes^n(p\,dq)\, f(q,p)\, dq\wedge dp$}$=A_n(q)\otimes^{n+1}\!dq$,
where $\otimes^{n}dq:=dq\otimes\dots\otimes dq\,$ $n$ times. Thus,  moments $A_n$
belong to the  vector space dual to the contravariant tensors of the type
$\beta_n=\beta_n(q)\otimes^{n}\!\partial_q$. Upon denoting the spatial derivative $\partial/\partial q$ by prime,
these tensors are given the following Lie algebra structure \cite{Ni55}
\begin{equation}
\left[\!\left[\alpha_m,\,\beta_n\right]\!\right]=
\big(\,n\,\beta_n(q)\,\alpha_m'(q)
-
m\,\alpha_m(q)\,\beta_n^{\,\prime}(q)\,\big)\otimes^{n+m-1}\!\partial_q=:
\textsf{\large ad}_{\alpha_m}\, \beta_n
\label{LieStruct-KM}
\end{equation}
so that the Kupershmidt-Manin Poisson bracket for moment dynamics is \cite{KuMa1978,GiHoTr2008}
\begin{equation}
\{F,H\}=
\left\langle A_{m+n-1},\,
\left[\!\!\left[\frac{\delta F}{\delta A_n},\frac{\delta H}{\delta A_m}\right]\!\!\right]
\right\rangle
\,,
\label{bracket-KM}
\end{equation}
where we sum over repeated indices. Thus, for a given moment Hamiltonian $H$, the Vlasov moment equations are
$\partial_t A_n
=
-\,\textsf{\large ad}^*_{{\delta H}/{\delta A_m}}\,A_{m+n-1}
$,
where the ${\sf ad}^*$ operator is defined by $\langle\, {\sf
ad}^*_{\beta_n} \,A_k,\,\alpha_{k-n+1}\,\rangle:= \langle\,
A_k,\,{\sf ad}_{\beta_n}\,\alpha_{k-n+1}\,\rangle$ and is given
explicitly as $ \textsf{ad}_{\beta_{n}}^{\ast}A_{k}=(\left(
k+1\right)  A_{k}\,\partial_{\,q\,} \beta_{n} +
n\,\beta_{n}\,\partial_{\,q\,} A_{k})\otimes^{k-n+2}dq$. In
the next section we use the following relation between the KM ${\sf
ad}^*\!$ operator and the canonical Poisson bracket
\[
\frac{\partial A_n}{\partial t}
=
\int\!p^n\,\frac{\partial f}{\partial t}\,{\rm d}p\,=-\sum_{m=0}^\infty\int\!p^n\left\{f,\,p^m\,\frac{\delta H}{\delta A_m}\right\}\, {\rm d}p\,=-\sum_{m=0}^\infty\,{\sf ad}^*_{\,{\delta H}/{\delta A_m}\,}A_{m+n-1}
\,,
\]
where we have used the chain rule formula ${\delta H}/{\delta
f}=p^m\,{\delta H}/{\delta A_m}$, that is ordinarily used to pass
from the Vlasov description to its corresponding moment formulation
\cite{Gi1981,GiHoTr2008}. For completeness, we report the following
basic fact underlying the Hamiltonian structure of 1D moment
dynamics
\begin{theorem}[Gibbons\cite{Gi1981}]\label{gibbonstheorem}
The operation of taking moments of Vlasov solutions is a Poisson
map.
\end{theorem}
This theorem was recently extended to any number of dimensions and to anisotropic interactions in $\cite{GiHoTr2008}$, where the Poisson map property is justified by the use of momentum maps.

%\comment{CT: remark taken out.}
%-----------------------------------------
\subsection{Geometric dissipation for moment dynamics and Darcy's Law}
Consider the following Lie algebra action on Vlasov densities:
$\beta_n \, \cdot
\,f:=\text{\large\pounds}_{X_{p^n\beta_n}}f=\big\{f,\,p^n\beta_n\big\}$
(no sum), which is naturally given by the canonical action of the
Hamiltonian function
$h(q,p)=p^n\beta_n(q)$.
The dual action defines the $\text{\Large$\star$}_{n}$ operator, given by
\begin{align}
\Big\langle f\,\text{\Large$\star$}_{n}\, g,\,\beta_n\Big\rangle
:=
\Big\langle f,\, \beta_n\, \cdot \,g\Big\rangle
&=
\int f\left\{g,p^n\beta_n\right\}{\rm d}q\,{\rm d}p
=
 \left\langle \int \big\{f, g \big\}\,p^n\,dp\,,\,\beta_n \right\rangle
 \,.
\label{stardef}
\end{align}
Consequently, the dissipative bracket for the moments is written as
\cite{HoPuTr2007-CR}
\begin{eqnarray*}
\frac{dF}{dt}= \{\!\{\,E\,,\,F\,\}\!\} &=& -\,\Bigg\langle \int\!
p^n\left\{\,\mu[f]\,, \frac{\delta E}{\delta f}\right\}dp,\,\int\!
p^n \left\{f\,,\frac{\delta F}{\delta f}\right\}dp \Bigg\rangle
 \,.
\end{eqnarray*}
Upon writing $\delta E/\delta f=p^k\beta_k$ and $\delta F/\delta
f=p^m\alpha_m$, the dissipative bracket becomes
\begin{equation}\label{diss-moments-bracket}
\{\!\{\,E\,,\,F\,\}\!\}=-\left\langle\textsf{\large ad}^*_{\beta_k}\, \widetilde{\mu}_{\,k+n-1},\,
\left(\textsf{\large ad}^*_{\alpha_m}A_{m+n-1}\right)^{\sharp}\,\right\rangle
\,,
\end{equation}
where $\widetilde{\mu}_s(q):=\int\! p^s
\mu[f]\,dp$. The purely dissipative dynamics for the moments is then given
by
\begin{equation}
\frac{\partial A_n}{\partial t}
=\textsf{\large ad}^*_{\gamma_m}A_{m+n-1}
\qquad\text{with}\qquad
\gamma_m:=\left(\textsf{\large ad}^*_{\beta_k}\, \widetilde{\mu}_{\,k+m-1}\right)^\sharp
\,.
\label{diss_moments}
\end{equation}
Here, the $\gamma_m$ are the tensor fields representing the `velocities' for the tensor densities $A_{m+n-1}$.

It turns out that the equation for $A_0$ carried along by the
velocity vector field $\gamma_1$ is exactly Darcy's law. Indeed,
upon dropping the sharp for simplicity and by truncating equation
\eqref{diss_moments} to consider only $m=1$, we obtain the equation
 $\partial_{\,t\,} \rho =\pounds_{\gamma_1\,}
\rho=\partial_{\,q\!}\left(\rho\, \gamma_1 \right)$, for the density
$\rho:=A_0$. If we assume that $E=E[\rho]$ to neglect inertial
effect and also that $\widetilde{\mu}_0=\mu[\rho]$ for simplicity,
then $\,\gamma_1=(\widetilde{\mu}_0 \,\partial\beta_0/\partial
q)^\sharp$ is recognised as the Darcy velocity. Therefore, the
density $\rho:=A_0$ evolves according to Darcy's law, namely
\begin{equation}
\frac{\partial \rho}{\partial t}
=
\frac{\partial}{\partial q}\!\left(\rho\,\mu[\rho]\,\frac{\partial}{\partial q}\frac{\delta E}{\delta
\rho}\right)
\,.
\label{Darcy-rho}
\end{equation}
Notice that no linear diffusion terms are present in the above
equation, since such terms can only arise from Brownian particle
motion that is not considered in this paper. Linear diffusion terms
can however be added {\it a posteriori} to account for stochastic
effects.
\begin{remark}[Local vs. nonlocal mobilities]
We notice that in all our discussions it is not necessary for the mobility quantity $\mu$  to be a functional of the dynamical variable. Indeed, even in the general formulation \eqref{FGdist} of double-bracket dissipation, the quantity $\mu[\kappa]$ must only satisfy the requirements for \eqref{FGdist} to be symmetric and positive definite. Thus, at the kinetic level, the phase-space mobility $\mu[f]$ in the dissipative Vlasov equation \eqref{diss_vlasov} can also be an ordinary function $\mu(f)$ (rather than a functional $\mu[f]$) of the Vlasov distribution $f(q,p)$ or even some fixed phase-space density $\mu(q,p)$. At the macroscopic level of Darcy's law, this would correspond respectively either to a function $\mu(\rho)$ of the macroscopic particle density $\rho(q)$ or to some fixed function $\mu(q)$ on physical space. In these cases, the double bracket Vlasov equation \eqref{diss_vlasov} simply produces Darcy's Law with complete absence of nonlocal terms. However, in what follows we shall consider the more general choice of a mobility functional, whose corresponding nonlocal effects have been investigated in \cite{HoPu2005}.
\end{remark}

\subsection{Vlasov derivation of dissipative moment equations}\label{Vlasov-darcy-derivation}
This section is devoted to clarify how one can pass from the dissipative Vlasov equation \eqref{diss_vlasov} to the double-bracket equations for the moments \eqref{diss_moments}. The starting point is to recognize that the double-bracket structure of equation \eqref{diss_vlasov} involves a sharp operation $(\,\cdot\,)^\sharp$ as follows
\begin{equation}
\frac{\partial f}{\partial t}
\,=\,
\left\{f\,,\left\{\mu[f]\,,\,\frac{\delta E}{\delta f}\,\right\}^{\sharp\,}
\right\}
\,,
\end{equation}
This sharp operation is evidently defined through an appropriate (co-)metric, and thus the most general form of the double-bracket Vlasov equation \eqref{diss_vlasov} is written as
\begin{equation}
\frac{\partial f}{\partial t}
\,=\,
\left\{f\,,\,K*\left\{\mu[f]\,,\,\frac{\delta E}{\delta f}\,\right\}
\right\}
\,.
\end{equation}
Here $K=K(q,p,q',p')$ is a symmetric positive-definite kernel, so that the sharp operator is given by the convolution $K*f=f^{\,\sharp}$ taking a density variable $f$ into a scalar function $f^\sharp$. The simplest choice of a $\delta$-function for $K$ evidently returns equation \eqref{diss_vlasov}.
In a more general case, one can consider a kernel $K$ that is analytic in the momentum coordinates, so that the double bracket structure
\begin{equation}
\{\!\{\,E\,,\,F\,\}\!\}
=
-\,\Bigg\langle\left\{\,f\,,\frac{\delta F}{\delta f}\right\} , \left\{\,\mu[f]\,, \frac{\delta E}{\delta f}\right\}^\sharp\Bigg\rangle
=
-\iint\!\left\{\,f\,,\frac{\delta F}{\delta f}\right\} \left(K* \left\{\,\mu[f]\,, \frac{\delta E}{\delta f}\right\}\right)dq\,dp
\end{equation}
becomes
\begin{equation}
\{\!\{\,E\,,\,F\,\}\!\}
%=
%-\,\Bigg\langle\left\{\,f\,,\frac{\delta F}{\delta f}\right\} , K*\left\{\,\mu[f]\,, \frac{\delta E}{\delta f}\right\}\Bigg\rangle
=-\sum_{n,m}\left\langle
\int \!p^{n\!}
\left\{
f,\frac{\delta F}{\delta f}
\right\}{\rm d}p
\ , \,K_{n,m}*
\int \!{p^{\prime\,}}^{m\!}
\left\{
\mu(f),\frac{\delta E}{\delta f}
\right\}{\rm d}p'
\right\rangle
\end{equation}
where we have used the Taylor expansion
\begin{equation}\label{Taylorexp}
K(q,p,q',p')=\sum_{n,m=0}^\infty p^n\,p^{\prime\,m\,} K_{n,m}(q,q')\,.
\end{equation}
\begin{assumption}
In what follows, we shall assume that the coefficients $K_{n,m}$ decay sufficiently fast so that the convergence of all the
necessary sums is guaranteed.
\end{assumption}
Upon expanding $\delta F/\delta f=\sum p^l\, \delta F/\delta A_l$ (and analogously for $\delta E/\delta f$) and integrating over the momentum variables, one obtains the double-bracket structure for the moments in the form
\[
\{\!\{\,E\,,\,F\,\}\!\}=-\sum_{n,l,k,m}\left\langle\textsf{\large ad}^*_{\beta_k}\, \widetilde{\mu}_{\,k+n-1\,},\, K_{n,l}*
\left(\textsf{\large ad}^*_{\alpha_m}A_{m+l-1}\right)\,\right\rangle
\,,
\]
which evidently recovers \eqref{diss-moments-bracket}.

Upon setting $k=m=0$ and $n=l=1$, the above bracket returns Darcy's law for $\rho=A_0$ in the more general form
\begin{equation}
\frac{\partial \rho(q,t)}{\partial t}
=
\frac{\partial}{\partial q}\left(\rho(q,t)\int K_{1,1}(q,q')\
\mu_\rho(q',t)\ \frac{\partial}{\partial q'}\frac{\delta E}{\delta \rho}\ {\rm d}q'\right)
\end{equation}
which recovers \eqref{Darcy-rho} when $K_{1,1}(q,q')=\delta(q-q')$. It is interesting to notice that, in the multi-dimensional treatment, the existence of the kernel $K_{1,1}$ allows for a \emph{mobility matrix}, rather than just a single-valued mobility functional.
Indeed, the quantity $K_{1,1}$ is always allowed to involve a symmetric positive matrix $\hat{K}_{1,1}$, as it can be seen by rewriting the Taylor expansion \eqref{Taylorexp} in the form
$K({\bf q,p,q',p'})=\sum {\bf p}^{\otimes\,n}\otimes {\bf p'}^{\otimes\,m}\!\contract \,\hat{K}_{n,m}({\bf q,q'})$,
where the symbol $\contract$ denotes tensor contraction and $\hat{K}_{n,m}$ is a covariant tensor field of order $n+m$.
Thus Darcy's law becomes, for example,
\begin{equation}
\frac{\partial \rho}{\partial t}
=
\operatorname{div}\left(\rho\left(
\mu_\rho\,\hat{K}_{1,1} \nabla\right)\frac{\delta E}{\delta \rho}\right)
=
\operatorname{div}\left(\rho\left(
\hat{\mu}_\rho \nabla\right)\frac{\delta E}{\delta \rho}\right)
\end{equation}
where we have introduced the matrix $\hat{K}_{1,1}$ so that $K_{1,1}({\bf q,q}')=\hat{K}_{1,1\,}\delta({\bf q-q}')$ and we have defined the mobility matrix $\hat{\mu}_\rho:=\mu_\rho\,\hat{K}_{1,1}$.
Physically, this corresponds to the case of isotropic particles moving through a non-isotropic medium.

The remainder of the paper concentrates on the case when the sharp operator $(\,\cdot\,)^\sharp$ is defined in terms of a delta-function kernel and the mobility $\mu_\rho$ is a real-valued functional of  $\rho$.

%\comment{CT: summary taken out}

\section{Dissipative dynamics with anisotropic interaction}\label{sec:Anisotropic}

\subsection{Purely dissipative Vlasov equation using GHK bracket}
Following GHK, we introduce a particle distribution which depends
not only on the position and momentum coordinates $q$ and $p$, but
also on an extra coordinate $g$ associated with, e.g.,
\emph{orientation} or some other order parameter. The coordinate $g$
belongs to the dual $\mathfrak{g}^*$ of the Lie algebra
$\mathfrak{g}$ of the order parameter group $\mathcal{G}$, which for
anisotropic interactions is $\mathcal{G}=SO(3)$. For an introduction
to this approach in the theory of complex fluids, see \cite{Ho2002}.
In what follows, we shall formulate the problem in the general
context and analyze the case of rotations separately. In the
non-dissipative case, the Vlasov equation is written in terms of a
Poisson bracket, which is the direct sum of the canonical
$(q,p)$-bracket and the Lie-Poisson bracket on the dual Lie algebra
$\mathfrak{g}^*$. Explicitly, this Poisson  bracket is written as
\eqref{PB1}.
%\begin{framed}\noindent
Like in the isotropic case, the non-dissipative Vlasov equation now
becomes
\[
\frac{\partial f}{\partial t} = -\left\{f,\frac{\delta H}{\partial
f}\right\}_{\!1}
%= -\, \widehat{X}_{\!\frac{\delta H}{\delta f}}(f)
\,,
\]
%where we have defined the vector field $\widehat{X}_h$ associated with the Hamiltonian function $h$ as
%\[
%\widehat{X}_h
%:=
%\frac{\partial h}{\partial p}\frac{\partial }{\partial q}
%-
%\frac{\partial h}{\partial q}\frac{\partial }{\partial p}
%+\left\langle{\rm ad}^*_{\frac{\partial h}{\partial g}}\,g,\,\frac{\partial }{\partial g}\right\rangle
%=
%X_h+
%\left\langle{\rm ad}^*_{\frac{\partial h}{\partial g}}\,g,\,\frac{\partial }{\partial g}\right\rangle
%\, .
%\]
Analogously to the isotropic case, the above Vlasov equation is
%a {\bfi characteristic equation} for evolution governed by the flow of the vector field $\widehat{X}_{{\delta H}/{\delta f}}$,
determined by the
%action of this vector field on the density $f$. One can identify $\widehat{X}_h$ with $h$ and define a
Lie-algebra action $h\cdot f:=\left\{f,h \right\}_{1}$ of
phase-space functions $h$ on phase-space distributions $f$. The
corresponding dual operation $(\star)$ is given by $\left(f\star k,
h \right) = \left(k,-\,h\cdot f \right) = \left(\,k, \,\left\{h,f
\right\}_{1} \,\right)=\left(\,\left\{f,k\right\}_{1},
\,h\,\right)$. Here we have denoted by $\left(\cdot,\cdot\right)$
the $L^2$ pairing between  functions and distributions on phase
space and the  equality $\left(\,k, \,\left\{h,f \right\}_{1}
\,\right)=\left(\,\left\{f,k\right\}_{1}, \,h\,\right)$ follows from
Leibniz rule for Poisson brackets. Thus, $f\star
k=\left\{f,k\right\}_{1}$.
%\end{framed}\noindent
%\begin{remark}
Upon applying the same arguments as in the previous Section~\ref{sec:Dissvlasov} and making use
of the general formulation of the dissipative bracket (\ref{bracket}), we
find the purely dissipative Vlasov equation in double-bracket form (anisotropic DBV equation):
\begin{equation}
\frac{\partial f}{\partial t}
=
\left\{f,\left\{\mu[f],\frac{\delta E}{\partial f}\right\}_{\!1}\right\}_{\!1}
\, ,
\label{Vlasov-diss1}
\end{equation}
where we have dropped the sharp symbol.
This equation  has exactly the same form as in (\ref{Vlasov-diss}), but now one replaces the
canonical Poisson bracket $\left\{\cdot\,,\cdot\right\}$ with
 the direct sum Poisson bracket
$\left\{\cdot\,,\cdot\right\}_1$ in  (\ref{PB1}). This formulation
can now be used to derive the double-bracket dissipative version of
the Vlasov equation for particles undergoing anisotropic
interaction.
%\end{remark}

\subsection{Dissipative moment dynamics in 1D}
An extension of the 1D moment dynamics hierarchy to include orientation dependence, may be obtained by following the same steps as in the previous section, beginning by introducing  the moment quantities $A_n(q,g)$ in \eqref{Smoluchowski-moments}.
One may find the entire hierarchy of equations for these moment quantities and then integrate over $g$ in order to find the equations for the mass density $\rho(q):=\int \!A_0(q,g)\,dg$ and the continuum charge density $G(q)=\int\! g\,A_0(q,g)\,dg$. Without the integration over $g$, such an approach would yield the Smoluchowski approximation for the density $A_0(q,g)$, usually denoted by $\rho(q,g)$. This approach is developed in Sec.~\ref{sec:Smoluchowski}, where the dynamics of $\rho(q,g)$
is presented explicitly.

In this section, we  extend the Kupershmidt-Manin approach as in GHK to generate the dynamics of moments with respect to both momentum $p$ and charge $g$. The main complication is that the Lie algebras of physical interest (such as $\mathfrak{so}(3)$) are not one-dimensional and in general are not commutative. Thus, in the general case one often introduces a multi-index notation as in \cite{Ku1987,GiHoTr2005}.
Rather, we introduce tensor powers $g^\nu:=g^{\,\otimes\,\nu}=g\otimes\dots\otimes g$. Then, the moments are expressed as
\begin{equation}
\label{Darcy-moments2}
A_{n,\nu}(q)\,:=\int p^n\,g^\nu\,f(q,p,g)\,dp\,dg
\,.
\end{equation}
Note that all the integrals with respect to $dp$ and $dg$ are taken in finite-dimensional spaces and are presumed to exist according to Assumptions \ref{assumption-finitedim} and \ref{assumption-existence}.
 This multi-dimensional treatment leads to
cumbersome calculations \cite{GiHoTr2008}. For the purposes of this
section, we are primarily interested in the equations for $\rho$ and
$G$; so we shall restrict our considerations to moments of the form
\makebox{$A_{n,0}=\int p^n\,f(q,p,g)\,dp\,dg$} and
\makebox{$A_{n,1}=\int p^n\,g\,f(q,p,g)\,dp\,dg$}. We write the
single particle Hamiltonian as $h=\delta H/\delta f=p^n
g^\nu\contract \delta H/\delta A_{n,\nu}=:p^n g^\nu
\contract\beta_{n,\nu}(q)$. Here $\contract$ denotes tensor
contraction between lower and upper indexes.  We restrict our
considerations to the case $\nu=0,1$. Consequently, we employ the
following assumption.

\medskip

%--------------------------------
%
\begin{assumption} \label{linassump}
The single-particle Hamiltonian $h=\delta H/\delta f$ is linear in
$g$ and can be expressed in 1D as $h(q,p,g)=
p^n\,\psi_n(q)+p^n\!\left\langle g,\,\overline\psi_n(q)
\right\rangle$, where $\psi_n(q)\in \mathbb{R}$ is a real scalar
function and $\overline\psi_n(q)\in \mathbb{R}\otimes\mathfrak{g}$
is a real Lie-algebra-valued function. This assumption will be used
throughout the rest of the paper.
\end{assumption}
%

%--------------------------------

\subsubsection{Dual Lie algebra action}
The action of $\beta_{n,\nu}$ on $f$ is defined as
$\beta_{n,\nu}\cdot f=\big\{p^n
g^{\nu\!}\contract\beta_{n,\nu},\,f\big\}_{\!1}$  (no sum), where
$\contract$ denotes tensor contraction between lower and upper
indexes. The dual of this action is denoted by $(\star_{n, \nu})$.
It may be  computed analogously to (\ref{stardef}) and found to be
\begin{align}\nonumber
f\,\,\text{\Large$\star$}_{n,\nu}\,k
\,=&\,
\iint  p^n g^\nu\,\big\{f,\,k\big\}_{\!1}\,{\rm d}p \,{\rm d}g\\
\nonumber
\,=&\,
\int g^{\nu+\sigma}\!\contract\,{\sf ad}_{\alpha_{m,\sigma}}^*A_{m+n-1} \,{\rm d}g
+
\int g^\nu
\left\langle g,\left[\frac{\partial A_{m+n}}{\partial g},\frac{\partial (g^\sigma\!\contract\alpha_{m,\sigma})}{\partial g}\right]\right\rangle{\rm d}g
\\
\,=&\,\,
{\sf ad}_{\alpha_{m,\sigma}}^*\int g^{\nu+\sigma} A_{m+n-1} \,{\rm d}g
+
\int g^\nu
\left\langle g,\left[\frac{\partial A_{m+n}}{\partial g}, \alpha_{m,1}\right]\right\rangle{\rm d}g
\, .
\label{darcy-moment-star}
\end{align}
Here, $k=p^m\,g^{\sigma\!}\contract\alpha_{m,\sigma}(q)$ (with $\sigma=0,1$) and we have used the definition of the moments $A_n(q,g)$ in \eqref{Smoluchowski-moments}.
Moreover, the notation ${\sf ad}^*_{\alpha_{m,\sigma}} A_{n,\sigma+\nu}$ above involves a tensor index contraction (arising from the contraction symbol in the second line of \eqref{darcy-moment-star}) between the upper and lower Lie algebra indexes. For example, in the simplest case one has ${\sf ad}^*_{\alpha_{0,1}} A_{0,1}=(A_{0,1})_a\,\partial_q(\alpha_{0,1})^a$.

Once the dual of the Lie algebra action is completely characterized, we can now write the double bracket structure for the moments $A_{n,\nu}$. This is done in the following discussion.

\subsubsection{Evolution equations}
Having characterized the dual Lie algebra action in \eqref{darcy-moment-star}, we
may write the evolution equation for an arbitrary functional $F$ in terms of the {\bfi dissipative bracket} as follows:
\begin{equation}
\dot{F}= \{\!\{\,F\,,\,E\,\}\!\}=-
\sum_{n,\nu=0}^\infty
\left\langle\!\!\!\left\langle \left(\mu[f]\,\,\text{\Large$\star$}_{n,\nu}\,\,\frac{\delta E}{\partial f}\right)^\sharp\!,\,f\,\,\text{\Large$\star$}_{n,\nu}\,\,\frac{\delta F}{\partial f} \right\rangle\!\!\!\right\rangle
\label{dissbracket1}
\end{equation}
where the pairing
$\left\langle\!\!\!\left\langle\,\cdot\,,\cdot\,\right\rangle\!\!\!\right\rangle$
is given by contraction of all upper and lower tensor indexes and
integration over the spatial coordinate $q$. In the one dimensional
case under consideration, we have explicitly
$f\,\,\text{\Large$\star$}_{n,\nu}\,{\delta F}/{\delta f} = \big(
f\,\,\text{\Large$\star$}_{n,\nu}\,{\delta F}/{\delta f} \big)_{i_1,
\ldots, i_\nu} \mathbf{e}^{i_1} \otimes \ldots \otimes
\mathbf{e}^{i_\nu}$, with $\left\{ \mathbf{e}^i \right\} $ being the
basis in the dual of Lie algebra.

We now consider the lowest order by setting $n=0,1$ and excluding the case $(n,\nu)=(1,1)$ (we have $\nu=0,1$ as before).  This is exactly the same truncation that is involved, on the Hamiltonian side, in the Vlasov moment equations of chromohydrodynamics \cite{GiHoKu1982,GiHoKu1983}. Thus, upon denoting $k=\delta F/\delta f$, we evaluate
\begin{align}\nonumber
f\star_{0,0}k&= {\sf
ad}^*_{\textstyle\alpha_{m,\sigma}\,}A_{m-1,\,\sigma}={\sf
ad}^*_{\textstyle\alpha_{m,0}\,}A_{m-1,\,0}+{\sf
ad}^*_{\textstyle\alpha_{m,1}\,}A_{m-1,\,1}
\\\nonumber
f\star_{0,1}k&=  {\sf
ad}^*_{\textstyle\alpha_{m,\sigma}\,}A_{m-1,\,\sigma+1}+ \int g
\left\langle g,\left[\frac{\partial A_{m}}{\partial g},
\alpha_{m,1}\right]\right\rangle{\rm d}g={\sf
ad}^*_{\textstyle\alpha_{m,0}\,}A_{m-1,\,1}+{\rm
ad}^*_{\textstyle\alpha_{m,1}} A_{m,1}
\\
f\star_{1,0}k&=  {\sf
ad}^*_{\textstyle\alpha_{m,\sigma}\,}A_{m,\,\sigma} = {\sf
ad}^*_{\textstyle\alpha_{m,0}\,}A_{m,\,0}+ {\sf
ad}^*_{\textstyle\alpha_{m,1}\,}A_{m,\,1} \label{stars}
\end{align}
where $k=p^m\,g^{\sigma\!}\contract\alpha_{m,\sigma}(q)$  and we
have truncated to consider only moments of the type $A_{m,0}$ and
$A_{m,1}$. Note that the notation for the Kupershmidt-Manin $\sf
ad^*$ operator differs from the notation for the coadjoint operator
$\rm ad^*$ associated to the Lie algebra $\mathfrak{g}$. At this
point, we consider the simplest case in which $m=0$, so that
%\begin{framed}\noindent
the first two relations of \eqref{stars} are  $f\star_{0,0}k=0$ and
$f\star_{0,1}k= {\rm ad}^*_{\alpha_{0,1}\,}A_{0,\,1}$, while the
third of  \eqref{stars} becomes
%\end{framed}\noindent
\begin{align*}
f\star_{1,0}k&=   {\sf ad}^*_{\textstyle\alpha_{0,0}\,}A_{0,\,0}+
{\sf ad}^*_{\textstyle\alpha_{0,1}\,}A_{0,\,1} =A_{0,\,0\
}\partial_{q\,}\alpha_{0,0}+(A_{0,\,1})_{a\,
}\partial_{q}(\alpha_{0,1})^a\,.
\end{align*}
The equation for the evolution of $F=A_{0,\lambda}$ is found from the bracket (\ref{dissbracket1}) in the truncated form
\begin{align*}
\dot{F}= \{\!\{\,F\,,\,E\,\}\!\} &=-\left\langle\!\!\!\left\langle
\left(\mu[f]\,\,\text{\Large$\star$}_{0,1}\,\,\frac{\delta
E}{\partial
f}\right)^\sharp\!,\,f\,\,\text{\Large$\star$}_{0,1}\,\,\frac{\delta
F}{\partial f} \right\rangle\!\!\!\right\rangle -
\left\langle\!\!\!\left\langle
\left(\mu[f]\,\,\text{\Large$\star$}_{1,0}\,\,\frac{\delta
E}{\partial
f}\right)^\sharp\!,\,f\,\,\text{\Large$\star$}_{1,0}\,\,\frac{\delta
F}{\partial f} \right\rangle\!\!\!\right\rangle
\\
&
=
-
\left\langle\!\!\!\left\langle \left(\mu_{0,\,0\, }\frac{\partial\beta_{0,0}}{\partial q}+\left\langle \mu_{0,1}\,,\frac{\partial\beta_{0,1}}{\partial q}\right\rangle\right)^\sharp\!,\,\left(A_{0,\,0\, }\frac{\partial\alpha_{0,0}}{\partial q}+\left\langle A_{0,\,1}\,,\frac{\partial\alpha_{0,1}}{\partial q}\right\rangle\right) \right\rangle\!\!\!\right\rangle
\\
&\quad -\left\langle\!\!\!\left\langle \left({\rm ad}^*_{\beta_{0,1}\,}\mu_{0,1}\right)^\sharp\!,\,{\rm ad}^*_{\alpha_{0,1}\,}A_{0,1} \right\rangle\!\!\!\right\rangle
\end{align*}
where the prime in the summation symbol stands for $n\neq\nu$ and we
have expanded $\delta E/\delta f=\beta_{0,0}+\langle
g,\beta_{0,1}\rangle={\delta E}/{\delta A_{0,0}}+\langle g,{\delta
E}/{\delta A_{0,1}}\rangle$. Also, we use the notation
$\mu_{n,\nu}(q)=\int p^n\,g^\sigma\mu[f]\,dp\,dg$. The sharp
operator in the above formulas involves an appropriate metric: all
the discussions from section $\ref{Vlasov-darcy-derivation}$
involved in the derivation of Darcy's law also hold in this context.

We now simplify the notation by defining the following dynamic
quantities: $\rho=\ A_{0,0}$ and $G= A_{0,1}$. Likewise, we define
the \emph{mobilities} as: $\mu_\rho= \mu_{0,0}$ and $\mu_G=
\mu_{0,1}$. In terms of these quantities, we obtain the following
result.

\begin{theorem}\label{momeqns-thm-simplified}
The 1D moment equations for $\rho$ and $G$ are given by
\begin{align}
\frac{\partial \rho}{\partial t}=&\,\,
\frac{\partial}{\partial q}\Bigg(\rho\,\,
%\overset{\Gamma_0:=\gamma_{10}^\sharp}{\overbrace{
\bigg(
\mu_\rho\,\, \frac{\partial}{\partial q}\frac{\delta E}{\delta \rho}
+
\bigg\langle \mu_G, \,\frac{\partial}{\partial q}\frac{\delta E}{\delta G}\bigg\rangle
\bigg)
%}}
\,\,
\Bigg)
\label{rhogen-simplified}
\\
\frac{\partial G}{\partial t}=&\,\,
\frac{\partial}{\partial q}\Bigg(G\left(\mu_\rho\,\, \frac{\partial}{\partial q}\frac{\delta E}{\delta \rho}+
\left\langle \mu_G, \,\frac{\partial}{\partial q}\frac{\delta E}{\delta G}\right\rangle\right) \Bigg)
+ {\rm ad}^*_{\left(\!{\rm ad}^*_{\frac{\delta E}{\delta G}} \mu_G\!\right)^{\!\!\sharp}}\,G
\label{Ggen-simplified}
\,.
\end{align}
\end{theorem}
Equations in this family (called Geometric Order Parameter
equations) were derived via a different approach in
\cite{HoPu2005,HoPu2006,HoPu2007}. These equations were also derived
by using a multi-index approach in \cite{HoPuTr2008}.
\begin{remark}[Higher dimensions]\label{3Dremark}
Gibbons' Theorem \ref{gibbonstheorem} has recently been extended to
account for more geometric structure in \cite{GiHoTr2008}, where the
authors also provide a general setting for the multi-dimensional
treatment. The higher dimensional treatment of the moment Poisson
structure provides a systematic machinery that enables us to write
 the double bracket moment equations in the full three
dimensional case. In 3D, the equations \eqref{rhogen-simplified} and
\eqref{Ggen-simplified} become
\begin{align}\nonumber
\frac{\partial \rho}{\partial t}=&\,\,
\operatorname{div}\Bigg(\rho\,\, \bigg( \mu_\rho\,\nabla\frac{\delta
E}{\delta \rho} + \bigg\langle \mu_G, \nabla\frac{\delta E}{\delta
G}\bigg\rangle \bigg) \,\, \Bigg)
\\
\frac{\partial G}{\partial t}=&\,\,
\operatorname{div}\Bigg(G\left(\mu_\rho\,\nabla\frac{\delta
E}{\delta \rho}+ \left\langle \mu_G, \nabla\frac{\delta E}{\delta
G}\right\rangle\right) \Bigg) + {\rm ad}^*_{\left(\!{\rm
ad}^*_{\frac{\delta E}{\delta G}}\, \mu_G\!\right)^{\!\sharp}}\,G
\,,
\end{align}
For the seek of simplicity, this paper considers only the one
dimensional case.
\end{remark}

\subsection{Singular solutions}
Equations (\ref{rhogen-simplified}) and (\ref{Ggen-simplified}) admit singular solutions. This means that the trajectory of a single fluid
particle is a solution of the problem and all the microscopic information
about the particles is preserved. We shall prove the following.
\begin{theorem}
Equations (\ref{rhogen-simplified}) and (\ref{Ggen-simplified}) admit
solutions of the form
\begin{align}
\rho(q,t)&=w_\rho(t)\,\,\delta(q-Q_\rho(t)) \,,\qquad\quad
G(q,t)=w_G(t)\,\,\delta(q-Q_G(t)) \label{singansatz}
\end{align}
where $\dot{w}_\rho=0$, $\dot{w}_G=\text{\rm ad}^*_{\gamma_0}\,w_G$
and
\begin{align*}
\dot{Q}_\rho&=
-\left(\mu_\rho\,\, \frac{\partial}{\partial q}\frac{\delta E}{\delta \rho}
+
\bigg\langle \mu_G, \,\frac{\partial}{\partial q}\frac{\delta E}{\delta G}\bigg\rangle\right)_{q=Q_\rho}
\qquad\quad
\dot{Q}_G=
-\left(\mu_\rho\,\, \frac{\partial}{\partial q}\frac{\delta E}{\delta \rho}
+
\bigg\langle \mu_G, \,\frac{\partial}{\partial q}\frac{\delta E}{\delta G}\bigg\rangle\right)_{q=Q_G}
\end{align*}
\end{theorem}
A proof of this theorem is easily obtained by taking the pairing of
 equations \eqref{rhogen-simplified} and \eqref{Ggen-simplified}
 with a couple of test functions $(\phi_\rho(q),\phi_G(q))$, integrating
 by parts and then substituting the singular solution anstatz. Again, we shall assume that all the necessary integrals exist since test functions $(\phi_\rho(q),\phi_G(q))$ are smooth, supposed to decay to zero sufficiently fast for large $q$  if necessary and $(\rho, G)$ together with the first derivatives are in  the generalized function space.

 A similar result applies for the Geometric Order Parameter (GOP)
equations investigated in \cite{HoPu2005,HoPu2006,HoPu2007}.

\subsection{An illustrative application: a filament of rod-like particles}
In this case $G={\bf m}(x)$, $\rm ad_\mathbf{v} \mathbf{w}=\mathbf{v} \times \mathbf{w}$ and $\rm ad^*_\mathbf{v} \mathbf{w}=-\mathbf{v} \times \mathbf{w}$, and the
Lie algebra pairing is represented by the dot product of vectors in $\mathbb{R}^3$. Therefore the equations are
\begin{align}
\frac{\partial \rho}{\partial t} &= \frac{\partial}{\partial x}
\Bigg( \rho\, \bigg( \mu_\rho\, \frac{\partial}{\partial
x}\frac{\delta E}{\delta \rho} +\, \boldsymbol\mu_{\bf
m}\!\cdot\frac{\partial}{\partial x}\frac{\delta E}{\delta \bf m}
\bigg) \, \Bigg) \label{rodrho}
\\
\frac{\partial {\bf m}\,}{\partial t} &= \frac{\partial}{\partial
x}\Bigg({\bf m}
%\underset{\Gamma:=\gamma_{1a}^\sharp\mathbf{e}_a}{\underbrace{
%\Big(
\left(\mu_\rho\frac{\partial}{\partial x}\frac{\delta E}{\delta \rho}
+
\boldsymbol\mu_{\bf m}
\cdot\frac{\partial}{\partial x}\frac{\delta E}{\delta {\bf m}}
\right)
\, \Bigg)
+
%\mu_{\bar{p}}\,\,
{\bf m}\times
\left(
\boldsymbol{\mu}_{\bf m}\times\frac{\delta E}{\delta \bf m}
\right)
\label{rodGilbert}
\end{align}
Note that equations for  density $\rho$   and orientation ${\bf m}$ have conservative parts (coming from the divergence of a flux). In addition, when $\boldsymbol{\mu}_{\bf m}=a{\bf m}$ for a constant $a$, the orientation ${\bf m}$ has precisely the dissipation term ${\bf m}\times{\bf m}\times \delta E/\delta \bf m$ introduced by Gilbert \cite{Gilbert1955}. Thus, we have derived the Gilbert dissipation term at the macroscopic level, starting from double-bracket dissipative terms in the kinetic theory description. As far as we know, this is the first time that the Gilbert dissipation term has been derived from a kinetic theory model.

%--------------------------------------------------------
%--------------------------------------------------------

\section{Smoluchowski approach\label{sec:Smoluchowski} }
We shall now turn our attention to the Smoluchowski approach for the description of the interaction of anisotropic particles. Usually, these particles are assumed to be rod-like, so their orientation can be described by a point on a two-dimensional sphere $S^2$ \cite{DoEd1988}. In this picture, one defines a distribution $\rho(\boldsymbol{q,\Omega},t)$ on the extended configuration space $\Bbb{R}^3\times S^2$, whose dynamics is usually regulated by the Smoluchowski equation
\begin{equation}
\frac{\partial \rho}{\partial t}=\operatorname{div}_{\boldsymbol{q}\!}\left(\rho\,\mu_T(\rho)\,\nabla_{\!\boldsymbol{q}}\frac{\delta E}{\delta \rho}\right)+\operatorname{div}_{\boldsymbol{\Omega}\!}\left(\rho\,\mu_O(\rho)\,\nabla_{\!\boldsymbol{\Omega}}\frac{\delta E}{\delta \rho}\right)
\label{oldSmoluchowski}
\end{equation}
where $\mu_T$ and $\mu_O$ are respectively the mobilities in the translational and orientational part and the rotational gradient operator $\nabla_{\!\boldsymbol{\Omega}}$ is expressed as $\nabla_{\!\boldsymbol{\Omega}}=\boldsymbol{\Omega}\times\partial/\partial\boldsymbol{\Omega}$. Upon combining the previous derivation of Darcy's law with the definition of the two different mobilities, one easily recognizes that the above equations can be obtained by following the same steps as in Darcy's law. Two different mobilities $\mu_T$ and $\mu_O$ involving different smoothing kernels are allowed by the particular Cartesian product structure of the configuration space $\Bbb{R}^3\times S^2$. The reference Vlasov equation for the above Smoluchowski equation is written as
\begin{align}
\frac{\partial f}{\partial t}
=
\left\{f\,,\,\left\{\mu_T[{f}],\frac{\delta E}{\delta
f}\right\}^\sharp_{T^*\mathbb{R}^3\!}
+
\left\{\mu_O[{f}],\frac{\delta
E}{\delta f}\right\}^\sharp_{T^*\!S^2\,}\right\}_{T^*(\mathbb{R}^3\times
S^2)}
%\\
%+ \left[\,f,\left[\chi_T(\bar{f}),\frac{\delta E}{\delta
%f}\right]_{T^*\mathbb{R}^3\!}+\left[\chi_O(\bar{f}),\frac{\delta
%E}{\delta f}\right]_{T^*\!S^2\,}\right]_{T^*\! S^2}
\label{split-DBf}
\end{align}
This principle could also be used to define two different mobilities $\boldsymbol{\mu}_{\bf m}^T$ and  $\boldsymbol{\mu}_{\bf m}^O$ in equation \eqref{rodGilbert}, so that $\boldsymbol{\mu}_{\bf m}^T$ would only appear in the expression of Darcy's velocity, while $\boldsymbol{\mu}_{\bf m}^O$ would appear in the Gilbert dissipation term involving the double cross product. For the sake of simplicity, we shall not deal with such a distinction between these two types of mobilities.

In what follows, we shall consider particles of arbitrary shape, for
which one needs the full group $SO(3)$ to define their orientation.
We shall work with the corresponding Lie algebra $\mathfrak{so}(3)$
to conform to our theory. This construction particularly suits the
dynamics of ferromagnetic particles. Following the Smoluchowski
approach, moments are  defined  as $A_n(q,g):=\!\int
p^n\,f(q,p,g)\,dp$. As in the Kupershmidt-Manin approach, these
moments are dual to $\beta_n(q,g)$, which are introduced by
expanding the Hamiltonian function $h(q,p,g)$ as
$h(q,p,g)=p^n\,\beta_n(q,g)$. Substitution of these forms yields a
{\bfi Lie algebra bracket} for the quantities $\beta_n$ given by $
\left[\!\left[\beta_n,\alpha_m\right]\!\right]_1=
\left[\!\left[\beta_n,\alpha_m\right]\!\right]+\left\langle
g,\left[\beta_n^{\,\prime},\alpha_m^{\,\prime}\right]\right\rangle$,
where prime denotes partial derivative with respect to $g$ and we
have used the same notation as in Section \ref{Kup-Man}. A
substantial difficulty arises in this moment hierarchy because the
above Lie algebra structure does not allow for subalgebras, as
happens for the moments treated in the previous sections. Thus,
there is no rigorous way of truncating the hierarchy, which instead
needs to be truncated by ad hoc physical arguments. However, in what
follows we notice that Smoluchowski equation (for $SO(3)$
orientations, or ferromagnetic particles) can be obtained by taking
the zero-th order moment $A_0(q,g)$ of the oriented DBV equation.

\subsection{A cold-plasma closure and its dissipative dynamics}
We shall apply the double bracket approach to the following
equations for the fluid momentum $m(q)=\int p\,f(q,p,g)\, {\rm
d}p\,{\rm d}g$  and the probability distribution $\rho(q,g)=\int\!
f(q,p,g) \, {\rm d}p=A_0(q,g)$ in the orientation
$g\in\mathfrak{g}^*$:
\begin{align}\nonumber
\frac{\partial m}{\partial t}+\frac{\delta H}{\delta
m}\frac{\partial m}{\partial q}+2m\frac{\partial}{\partial
q}\frac{\delta H}{\delta m}&=-\int\!\rho\,\frac{\partial}{\partial
q} \frac{\delta H}{\delta \rho}\,{\rm d}g\,, \qquad\
\frac{\partial\rho}{\partial t}+\frac{\partial}{\partial
q}\left(\rho\,\frac{\delta H}{\delta m}\right)=-\left\langle
g,\left[\frac{\partial \rho}{\partial g},\frac{\partial}{\partial
g}\frac{\delta H}{\delta \rho}\right]\right\rangle \label{newmodel}
\end{align}
These equations arise from the Vlasov cold-plasma solution
\[
f(q,p,g)=\rho(q,g)\ \delta\!\left(p-\frac{m(q)}{\int\!\rho(q,g)\,{\rm d}g}\right)
\]
and their Hamiltonian structure is directly inherited from the
GHK-Vlasov bracket \cite{GiHoKu1983} $ \{F,H\}=\int\!
f\,\big\{{\delta F}/{\delta f},{\delta H}/{\delta f}\big\}_1{\rm
d}q\,{\rm d}p\,{\rm d}g\,, $ by direct substitution of the chain
rule formula ${\delta F}/{\delta f}={\delta F}/{\delta \rho}+p\
{\delta F}/{\delta m}$ for generic functionals $F$ and $H$.

In this picture, the {\bfi Lie algebra action} of the variable
$\left(\varphi(q,g),u(q)\right)\in
C^\infty(Q\times\mathfrak{g}^*)\times\mathfrak{X}(Q)$ is given by
$(\varphi,u)\cdot  f=\left\{f,\varphi+p\,u\right\}_1$ and its {\bfi
dual action} is given by
\begin{equation*}
\Big\langle f\,\text{\Large$\star$}\, k,\, (\varphi,u)\Big\rangle:=
- \Big\langle k,\, (\varphi,u)\cdot f\Big\rangle =
 \left\langle \left(\int \,\{f, k\}_{1}\,dp,\,\int p\,\{f, k\}_{1}\,dp\,dg\right),\,(\varphi,u) \right\rangle
 \,,
\end{equation*}
where the last pairing $\langle \,\cdot\,,\, \cdot\,\rangle:
(C^\infty(Q\times\mathfrak{g}^*)\times\mathfrak{X}(Q))\times
(C^\infty(Q\times\mathfrak{g}^*)\times\mathfrak{X}(Q))^*\to\mathbb{R}$
between fluid and particle variables is defined as the sum
%\label{fluid-particle-pair}
$ \big\langle (\rho, m),(\phi,u) \big\rangle = \int
\rho(q,g)\,\phi(q,g) \ dq\,dg + \int m(q)\,u(q)\,dq$.
More
explicitly, the {\bfi star operator} is defined for
$k(q,p,g)=\psi(q,g)+pv(q)$ as
\begin{align}
f\,\,\text{\Large$\star$}\,k
\,=&\,
-\left(\frac{\partial (\rho v)}{\partial q}+\left\langle g,\left[\frac{\partial \rho}{\partial g},\frac{\partial\psi}{\partial g}\right]\right\rangle\,,\ v\frac{\partial m}{\partial q}+2m\frac{\partial v}{\partial q}+\int\!\rho\,\frac{\partial\psi}{\partial q}\,{\rm d}g\right)
 \, .
 \label{smolstar}
\end{align}
At this point, we are ready to  introduce the {\bfi dissipative bracket} by
\begin{multline}\label{dissbracketSmol}
\dot{F}=\{\{F,E\}\}=-\left\langle\left(
\mu[f]\,\text{\large$\star$}\,\frac{\delta E}{\partial
f}\right)^\sharp,\,f\,\text{\large$\star$}\,\frac{\delta F}{\partial
f} \right\rangle
\\
=-\left\langle\left( \left\langle g,\left[\frac{\partial
\mu_\rho}{\partial g},\frac{\partial}{\partial g}\frac{\delta
E}{\partial
\rho}\right]\right\rangle,\int\!\mu_\rho\,\frac{\partial}{\partial
q}\frac{\delta E}{\partial \rho}\,{\rm d}g\right)^\sharp , \left(
\left\langle g,\left[\frac{\partial \rho}{\partial
g},\frac{\partial}{\partial g}\frac{\delta F}{\partial
\rho}\right]\right\rangle,\int\!\rho\,\frac{\partial}{\partial
q}\frac{\delta F}{\partial \rho}\,{\rm d}g\right) \right\rangle
\end{multline}
where $\mu_\rho=\int\mu\,dp$. Here we have restricted to consider only functionals of the probability distribution $\rho(q,g)$, so that $\delta E/\delta f=\delta E/\delta \rho$ and analogously for $F$ (no inertial effects).
By using this evolution equation for an arbitrary functional $F$, the rate of change for zero-th moment  $\rho(q,g)=\int f(q,p,g) \, {\rm d}p$ is found to be
\begin{equation}\label{newsmol}
\frac{\partial \rho}{\partial
t}=\frac{\partial}{\partial q}\left(\rho\int\mu_\rho\,\frac{\partial}{\partial q}\frac{\delta
E}{\delta \rho}\ {\rm d}g\right) +
\left\{\rho,\left\{\mu_\rho,\frac{\delta E}{\delta
\rho}\right\}_{\!\mathfrak{g}^*\!} \right\}_{\!\mathfrak{g}^*} \, ,
\end{equation}
where we have introduced the notation
$\{\cdot,\,\cdot\}_{\mathfrak{g}^*}$ for the Lie-Poisson bracket $
\left\{\varphi,\,\psi\right\}_{\mathfrak{g}^*}:= \left\langle
g,\left[\partial_{\,g\,} \varphi\,,\partial_{\,g\,}
\psi\right]\right\rangle$. The novel feature of equation
\eqref{newsmol} is the integral appearing in the expression of
Darcy's velocity (first term of the right hand side). While ordinary
Smoluchowski equations treat the translational and orientational
coordinates on the same level, equation \eqref{newsmol} treats these
coordinates from two very different perspectives, while retaining
the Yang-Mills charge nature of the orientation coordinate $g$. This
reflects in the appearance of the integral over the orientation $g$
that produces a velocity vector field defined on the manifold $Q$.
Such an integral is not present in conventional Smoluchowski
approaches.
\begin{remark}[Higher dimensions]
Analogously to remark \ref{3Dremark}, we can also write the three
dimensional version of equation \eqref{newsmol}. This reads as
\begin{equation*}
\frac{\partial \rho}{\partial
t}=\operatorname{div}\!\left(\rho\int\mu_\rho\,\nabla\frac{\delta
E}{\delta \rho}\ {\rm d}\mu\right) +
\left\{\rho,\left\{\mu_\rho,\frac{\delta E}{\delta
\rho}\right\}_{\!\mathfrak{g}^*\!} \right\}_{\!\mathfrak{g}^*}
\end{equation*}
\end{remark}

\subsection{Singular solutions}
Both conventional Smoluchowski equation (\ref{oldSmoluchowski}) and
equation \eqref{newsmol} possess delta-function solutions of the
type $\rho(q,g)=\Gamma\, \delta(q-Q(t))\,\delta(g-G(t))$, where
$\Gamma$ is a constant. This is the usual Klimontovich single
particle solution. It is an open question whether such solutions
emerge spontaneously. Another class of singular solutions is the
following
\begin{equation}\label{newpp-soln}
\rho(q,g)=w(g,t)\, \delta(q-Q(t)).
\end{equation}
This is a point particle solution that supports a kinetic equation for its own probability distribution $w(g,t)$ in the space ${\rm Den}(\mathfrak{g}^*)$ of densities in the orientational degree of freedom $g\in\mathfrak{g}^*$. In this case, the dynamical equations are
\[
\dot{Q}=-\left.\mu_\rho\,\frac{\partial }{\partial q}\frac{\delta E}{\delta \rho}\right|_{q=Q}\,,\qquad
\frac{\partial w}{\partial t}=\left.\left\{w,\left\{ \mu_\rho,\frac{\delta E}{\delta \rho}\right\}_{\!\mathfrak{g}^*}\right\}_{\!\mathfrak{g}^*}\right|_{q=Q}
\]
This is a coupled system for the ODE regulating the particle trajectory $Q(t)$ and the PDE governing the orientational probability $w(g,t)$. To our knowledge, this class of singular solutions is new.

\section{Summary and outlook}

The double-bracket Vlasov moment dynamics discussed here has provided an alternative to both the variational-geometric approach of \cite{HoPu2007} and the Smoluchowski approach  reviewed in \cite{Co2005}. These are early days in this study of the benefits afforded by the double-bracket approach to Vlasov moment dynamics. However, the derivations of the Darcy law in (\ref{Darcy-rho}) and the Gilbert dissipation term in (\ref{rodGilbert}) by this approach lends hope that this direction will provide the systematic derivations needed for modern technology of macroscopic models for microscopic processes involving interactions of particles that depend on their relative orientations. Although some of these formulas may look daunting, they possess an internal consistency and systematic derivation that we believe is worth pursuing further. Our next steps will be the following:
1) perform the analysis of the mobility functionals in kinetic space $\mu[f]$ as well as the mobilities for each particular geometric quantity $\mu_\rho$, $\mu_G$, \emph{etc.};
2) determine the conditions for the emergence of weak solutions (singularities) in the macroscopic (averaged) equations.

\subsection*{Acknowledgments} The authors were partially supported by NSF grants NSF-DMS-05377891 and NSF-DMS-09087551. The work of
DDH was also partially supported  by the Royal Society of London Wolfson Research Merit Award.

%%%%%%%%%%%%%%%%%%%%%
\bibliographystyle{unsrt}
%%\comment{Rearrange the bibliography in alphabetical order}
%\small

%%%%%%%%%%%%%%%%%%%%%}

\end{document}